\documentclass[aps,prl,twocolumn,superscriptaddress,showpacs]{revtex4}

\usepackage{graphicx}
\usepackage{dcolumn}
\usepackage{bm}

\begin{document}


\title{First access to $\beta$ half-lives approaching the r-process path near $N$=126}



\author{T.~Kurtukian-Nieto}
\email[]{kurtukia@cenbg.in2p3.fr}

\author{J.~Benlliure}
\affiliation{Universidad de Santiago de Compostela, E-15782 Santiago de Compostela, Spain}

\author{L.~Audouin}
\affiliation{Institut de Physique Nucl\'eaire, F-91406 Orsay Cedex, France}

\author{F.~Becker}
\affiliation{Gesellschaft f\"ur Schwerionenforschung mbH (GSI), Planckstr. 1, D-64291 Darmstadt, Germany}

\author{B.~Blank}
\affiliation{CENBG -
Universit\'e Bordeaux 1 - UMR 5797 CNRS/IN2P3, Chemin du Solarium, BP 120, 33175 Gradignan Cedex, France}

\author{I.N.~Borzov}
\altaffiliation[Present address: ] {IPPE, Bondarenko Sq.1, 249033,
Obninsk, Russia}

\affiliation{Gesellschaft f\"ur Schwerionenforschung mbH (GSI), Planckstr. 1, D-64291 Darmstadt, Germany}

\author{E.~Casarejos}
\affiliation{Universidad de Santiago de Compostela, E-15782 Santiago de Compostela, Spain}

\author{M.~Fern\'andez-Ord\'o\~nez}

\affiliation{Universidad de Santiago de Compostela,
E-15782 Santiago de Compostela, Spain}

\author{J.~Giovinazzo}
\affiliation{CENBG -
Universit\'e Bordeaux 1 - UMR 5797 CNRS/IN2P3, Chemin du Solarium, BP 120, 33175 Gradignan Cedex, France}

\author{D.~Henzlova}

\affiliation{Gesellschaft f\"ur Schwerionenforschung mbH (GSI),
Planckstr. 1, D-64291 Darmstadt, Germany}

\author{B.~Jurado}

\affiliation{Gesellschaft f\"ur Schwerionenforschung mbH (GSI), Planckstr. 1, D-64291 Darmstadt, Germany}

\author{K.~Langanke}
\affiliation{Gesellschaft f\"ur Schwerionenforschung mbH (GSI),
Planckstr. 1, D-64291 Darmstadt, Germany} \affiliation{Institut
f\"ur Kernphysik, TU Darmstadt, Schlossgartenstr. 9, D-64289
Darmstadt, Germany}

\author{G.~Mart\'{\i}nez-Pinedo}
\affiliation{Gesellschaft f\"ur Schwerionenforschung mbH (GSI), Planckstr. 1, D-64291 Darmstadt, Germany}

\author{J.~Pereira}

\affiliation{Universidad de Santiago de Compostela, E-15782
Santiago de Compostela, Spain}

\author{F.~Rejmund}
\affiliation{GANIL, BP55025, F-14076 CAEN Cedex 5, France}

\author{K.-H.~Schmidt}
\affiliation{Gesellschaft f\"ur Schwerionenforschung mbH (GSI), Planckstr. 1, D-64291 Darmstadt, Germany}

\author{O.~Yordanov}

\affiliation{Gesellschaft f\"ur Schwerionenforschung mbH (GSI), Planckstr. 1, D-64291 Darmstadt, Germany}

\date{\today}

\begin{abstract}
Heavy neutron-rich nuclei close to {\it N}=126 were produced by
fragmentation of a 1 {\it A} GeV $^{208}$Pb beam at the FRS at
GSI. The $\beta$-decay half-lives of 8 nuclides have been
determined. The comparison of the data with model calculations
including an approach based on the self-consistent ground-state
description and continuum QRPA considering the  Gamow-Teller and
first-forbidden decays provide a first indication on the
importance of first-forbidden transitions around {\it A}=195. The
measured data indicate that the matter flow in the r-process to
heavier fissioning nuclei is faster than previously expected.

\end{abstract}


\maketitle


The investigation of heavy neutron-rich nuclei in the laboratory
has been a challenging problem during the last decades. In
particular, the $\beta$-decay half-lives of nuclei on the
astrophysical r-process path are of primary importance for the
full understanding of this process \cite{cowan91}. They define how
rapidly the heaviest nuclei are synthesized during the r-process,
thus determining the strength of fission cycling \cite{cameron03}.
Moreover, these half-lives define the matter-flow bottlenecks,
which largely affect the r-process abundance pattern. However, the
scarce experimental data on r-process nuclei, along with the large
discrepancies between theoretical predictions which differ by
orders of magnitude, contribute to the difficulties in
understanding the astrophysical r-process. This is particularly
true for the $N \approx 126$ region far below the doubly magic
$^{208}$Pb, which remains a practically unexplored territory. The
half-lives of these nuclei determine the relative $A = 195$
r-process peak abundances as well as the matter flow to even
heavier nuclei.

In general there exist two complementary methods for providing and
for separating the nuclei of interest: the in-flight \cite{morrissey}
technique and the isotope-separation on-line (ISOL) technique \cite{duppen}.
Both methods have extensively been used for $\beta$-decay studies of
r-process-relevant nuclei at different regions of the chart of the nuclides
(e.g. around $^{78}$Ni \cite{hosmer} and $^{130}$Cd \cite{kratz}). The
ISOL technique suffers from specific difficulties for heavy neutron-rich nuclei.
Elements between Hf ($Z = 72$) and Pt ($Z = 78$) are difficult to extract from
the production target due to their low volatility, and for neutron-rich isotopes
of the adjacent heavier elements, the ISOL technique suffers from an overwhelming
isobaric contamination of isobars with higher $Z$, which are produced with much
higher cross sections. Background suppression was achieved for $^{218}$Bi
\cite{witte} as a peculiar case, by combining resonant-laser ionization and the
specific condition that the isotope of interest is longer lived than the unwanted
species. But these conditions are not generally applicable.

The in-flight separation, which is used in the present work, is
universal, offering the advantage of fast ($< 1 \mu s$), highly
efficient, chemistry-independent separation. By tuning the
magnetic fields of the separator, it is possible to select the
desired region on the chart of nuclides and to perform experiments
focusing either on one nucleus or on a `cocktail' with a
controlled background. Contaminants due to the small fraction of
inevitable secondary reactions, produced during the stopping
process, can be eliminated by a further range selection at the
implantation stage. As a consequence, the clean particle
identification for each implanted nucleus permits the direct
correlation of an individual fragment with its subsequent $\beta$
decay on an event-by-event basis. Important progress has been
obtained during the last years profiting from relativistic
heavy-ion beams, which made it possible to explore reaction
mechanisms leading to the production of heavy neutron-rich nuclei.
Cold-fragmentation reactions \cite{cold} opened a field not
accessible by fission and form the basis for the achievement of
the present work.

The experiment, which aimed at exploring the production of heavy neutron-rich nuclei close to
the neutron closed shell $N = 126$ and for measuring their $\beta$ half-lives, was performed at
the fragment separator FRS \cite{frs} at the Gesellschaft f\"ur Schwerionenforschung (GSI) in
Darmstadt, Germany. A $^{208}$Pb primary beam of 1 $A$~GeV, delivered by the SIS18 heavy-ion
synchrotron, was directed to a beryllium target at the entrance of the FRS. The reaction residues
were identified by determining both their atomic number $Z$ and their mass-over-charge ratio $A/Z$
by means of the measurements of the energy loss, the magnetic rigidities, and the time of flight (ToF).

\begin{figure}
\begin{center}
\includegraphics[width=\linewidth]{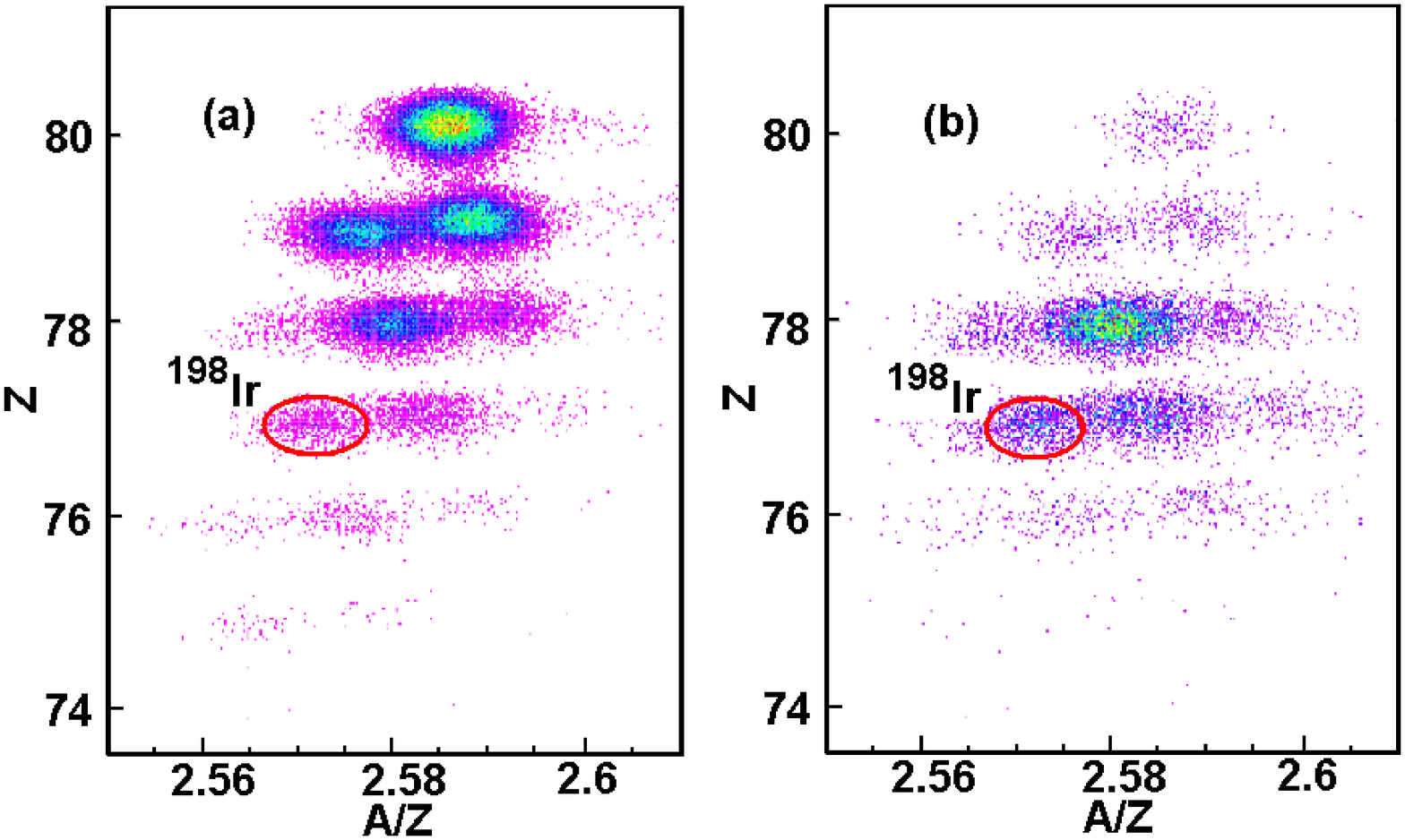}\\[3mm]
\includegraphics[width=0.8\linewidth]{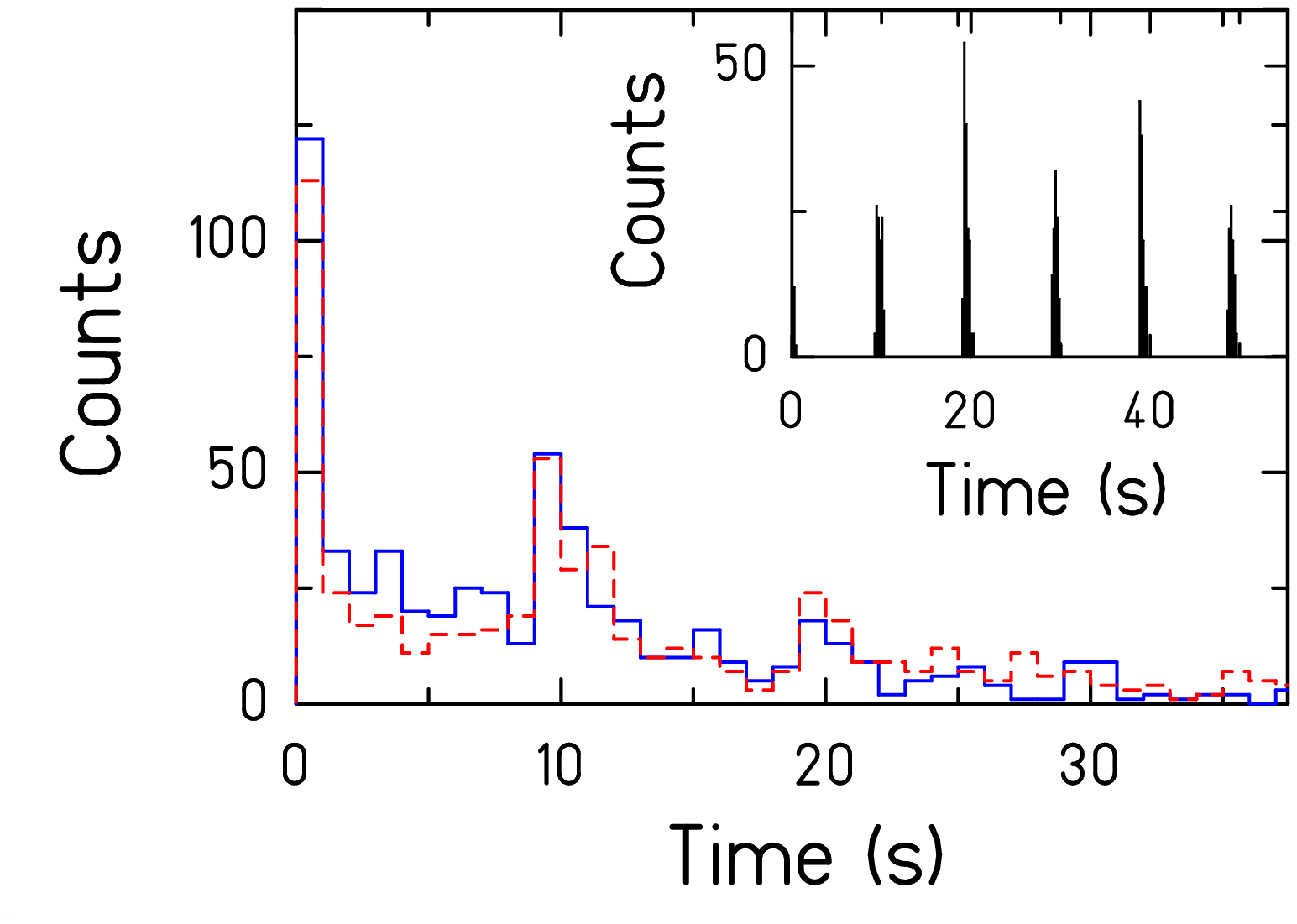}\\
\end{center}
\vspace{-4mm}
\caption{(Color online). Top: Particle
identification plots corresponding to a FRS setting optimized to
transmit $^{198}$Ir with a mono-energetic degrader. The figure
shows the total production yield at the end of the FRS (a) and the
ions which were implanted in the 1mm-thick DSSD (b). Bottom:
Time-difference spectra between the $^{195}$Re nuclei implanted
and the first $\beta$-like event detected in the same strip, in
forward- (solid line) and backward-time (dashed line). In the
inset, the time structure of the beam is shown. } \label{fig1}
\end{figure}

After leaving the FRS, the ions were slowed down to a few tens of
$A$~MeV in an aluminum plate and then implanted in an active
stopper: a highly pixelated double-sided 1mm thick Si detector
array (DSSD) with a surface of $20 \times 5$~cm$^2$, covering the
final image plane of the FRS. The use of a mono-energetic degrader
at the FRS provided a large horizontal dispersion and a narrow
range distribution of the fragments in the active stopper
\cite{degrader,degrader2}. The horizontal dispersion allowed
taking advantage of the high pixelation of the active catcher to
avoid multiple implantations of nuclei in the same pixel. The
narrow range distribution of the fragments increased the
implantation efficiency and allowed to implant several nuclides in
a thin active stopper ($< 1$~mm) at the same time. The
identification plot of the nuclides leaving the FRS for a magnetic
setting centred on  $^{198}$Ir is shown on the left upper part of
Fig. 1. The corresponding plot of the implanted nuclei in the
right upper part of Fig. 1 demonstrates the additional selection
by the implantation condition.

The half-lives of the nuclei were determined from position and time correlations between the
implantation of the fully identified nuclei and the first $\beta$-like event observed after
implantation, shown in the lower part of Fig. 1 for the case of $^{195}$Re. We observed a
beam-induced background in the recorded time-correlation spectra, originating from the time
structure of the previously implanted nuclei and the presence of $\delta$ or atomic electrons
generated during the beam pulse. The time structure of the beam, which is given by the periodic
operation cycles of the synchrotron, is shown in the inset in the lower part of Fig. 1. In order
to disentangle the background from the real events, the time differences between a given
implantation and the previous $\beta$-like event was accumulated, shown in Fig. 1 by the
dashed line. The ratio between the forward- and backward-time correlations, which decreases
with time, contains the information on the `true' fragment-$\beta$ correlations.

\begin{figure}[t]
\includegraphics[width=0.83\linewidth]{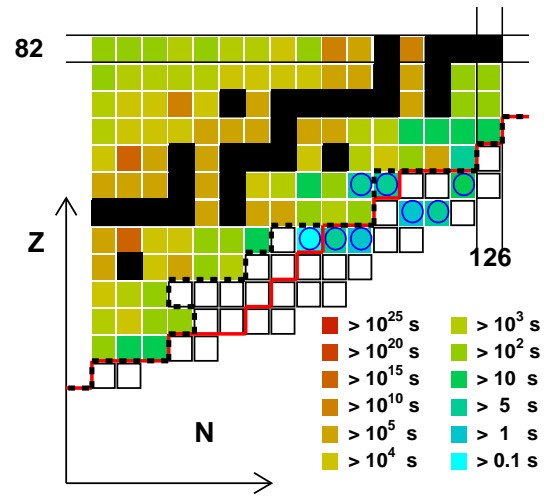}
\caption{(Color). Chart of the nuclides around $N$ = 126. The
  color scale
  indicates the known $\beta$ half-lives \cite{exp_beta}, while the
  circles correspond to the experimental values determined in this
  work. Empty boxes mark nuclei, identified in the
  present experiment, with no experimental half-life information
  available. The solid line denotes the limit of previously known
  nuclides and dashed line the limits of the known half-lives.}
\label{chart_of_nuclides}
\end{figure}

In order to cope with the complex time structure in the present data, we developed a new
method \cite{TKN_NIM} to determine the $\beta$-decay half-lives: The experimental data were
fitted with a numerical function (rather than an analytical function), obtained from Monte-Carlo
simulations of time correlations between implantations and $\beta$-decay events, with the conditions
found in the experiment. We performed sets of simulations with given values of $\beta$ efficiency
and lifetime, and we calculated the $\chi^2$ from the measured and simulated ratios of the spectra
of time correlations in forward and backward time for each set of simulations.  This new analysis
method solves a problem often arising in in-flight secondary-beam experiments when using
conventional analysis tools, if the primary beam is delivered by a synchrotron.

In Table I we list the half-lives determined in this work with their uncertainties. Figure 2 shows
the region of the chart of nuclides covered by the present work. The nuclear half-lives are indicated
by the color scale, and the circles represent the half-lives measured in this work. We were able
to synthesize more than 190 heavy neutron-rich nuclei `south' of lead, 25 of which were identified
for the first time.

\begin{table}
\label{tabla}
\caption{\textit{Measured $\beta$-decay half-lives compared with different model calculations.
}}
\begin{center}
\begin{ruledtabular}
\begin{tabular}{lrcrr}

Nuclide        & $T_{1/2}$         &  DF3+QRPA   & Gr. Th.  & FRDM  \\
               &   (s)             &   (s)      &   (s) \cite{gt2}    &  (s) \cite{MPK02}\\

\hline
$^{202}$Ir   &$ 11^{+3}_{-3}$    &  9.8       & 8.5     & 68.4  \\
$^{199}$Ir   &$ 6^{+5}_{-4}$     &  46.7       & 96.6    & 370.6 \\
$^{198}$Ir   &$ 8^{+2}_{-2}$     &  19.1      & ---     & 377.1 \\
$^{200}$Os   &$ 6^{+4}_{-3}$     &  6.9      & 16      & 187.1 \\
$^{199}$Os   &$ 5^{+4}_{-2}$     &  6.6      & 17.2    & 106.8 \\
$^{196}$Re   &$ 3^{+1}_{-2}$     &  1.4       & 5.1     &  3.6  \\
$^{195}$Re   &$ 6^{+1}_{-1}$     &  8.5       & 10.3    &  3.3  \\
$^{194}$Re   &$1^{+0.5}_{-0.5}$  &  2.1      & 16.1    & 70.8  \\

\end{tabular}

 \end{ruledtabular}
 \end{center}
\end{table}

The data of the new experiment do not extend to the extremely proton-deficient $N = 126$ nuclei on
the r-process path ($Z \approx 65 - 72$).  Therefore, r-process simulations still rely on theoretical
models for the $\beta$-decay half-lives. These calculations have been done globally, i.e. for all
r-process nuclei within the same framework, based on the Quasi-Particle Random Phase Approximation
(QRPA) on top of macroscopic-microscopic mass models like the Finite Range Droplet Model (FRDM)
\cite{Moeller97} or the Extended Thomas Fermi with Strutinski Integral (ETFSI) approach \cite{Borzov01},
while Hartree-Fock-Bogoliubov (HFB) \cite{Engel99} and shell model \cite{rmp} calculations have
been reported for the $N=126$ waiting point nuclei.  However, the later two calculations considered
only allowed Gamow-Teller (GT) transitions, while the work of Borzov \cite{B03}, based on the
continuum QRPA approach on top of mean-field ground states derived within the density-functional
(DF) theory, includes the GT and first-forbidden (FF) transitions consistently within the same
microscopic approach.  This study gives strong evidence for important contributions of FF
transitions to the total half-lives arising from $\nu 3p_{1/2}\rightarrow\pi 3d_{3/2}$ and
$\nu 1i_{13/2}\rightarrow\pi 2h_{11/2}$ neutron-proton transitions and hence implies pure GT
estimates of the half-lives of the $N=126$ waiting points insufficient. As a consequence, FF
transitions, however, derived within the statistical gross theory, have recently been added
to the FRDM/QRPA approach (noted FRDM in the following) \cite{MPK02}. Importantly the theoretical
models predict noticeably different half-lives for the $N=126$ waiting point nuclei making these
input data for r-process simulations quite uncertain.

Although the nuclei studied here are not located on the r-process path, they are close enough to
the $N=126$ waiting points to serve as stringent tests for the theoretical predictions. In Table I
we compare our measured half-lives with the prediction of the FRDM approach \cite{MPK02}, the
statistical 'gross theory' (Gr. Th.) \cite{gt2} and a new microscopic calculation (DF3+QRPA)
based on the Fayans energy-density functional. The later  follows the half-life calculations for
$N \sim 126$ nuclei as described in \cite{B03}. However, the strength functions for GT and FF
transitions are calculated with no energy-dependent smearing. We stress that the DF3 functional
used here describes the $Q_\beta$ values of neutron-rich nuclei quite well \cite{B03}. Furthermore,
the same approach has been proven to reproduce the measured half-lives of spherical nuclei in
different regions of the nuclear chart ($Z=20-30,N \approx 50$ \cite{B03}, $Z=42-49, N \approx
82$ \cite{B07}) quite well.

\begin{figure*}
\begin{center}
\includegraphics[width=0.7\linewidth]{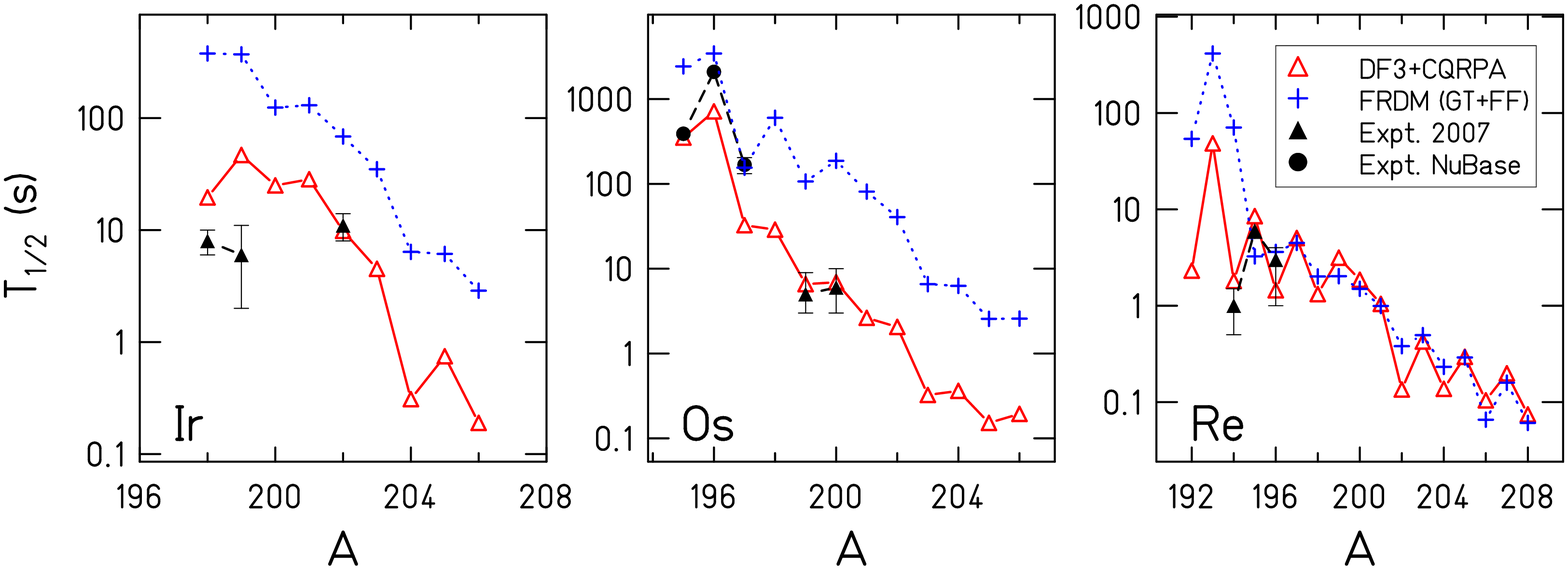}\\
\end{center}
\vspace{-6mm}
\caption{(Color online). The total $\beta$-decay
half-lives for Re, Ir and Os isotopic
  chains calculated within DF3+CQRPA in comparison with the ones
  from FRDM(GT+FF) \cite{MPK02}, both with Gamow-Teller plus
  first-forbidden transitions. Filled triangles correspond to the
  experimental values measured in this work, circles (Os isotopes) to
  half-lives taken from literature \cite{exp_beta}. }
\label{T_theor}
\end{figure*}

As can be seen in Table I, our DF3+QRPA model gives also a good account of the present half-lives,
with the one exception of $^{199}$Ir. The gross theory parametrization describes the data fairly well.
On the other hand, the FRDM half-lives are for most nuclei more than one order of magnitude longer than
the data. Fig. 3 compares half-lives from the FRDM and DF3+QRPA models for the Ir, Os and Re
chains, including the $N=126$ isotopes, and with data if available.  For the odd-Z Re isotopes
both models yield quite similar half-lives for the neutron-rich nuclei.  However, both for the
even-Z Os isotopes and odd-Z Ir isotopes the FRDM predicts substantially longer half-lives than
the DF3+QRPA calculations and the available data. The consistently good agreement of the results
of the spherical DF3+QRPA model with the presently reported half-lives, and with those in other
regions of the nuclear chart, places some confidence in applying this approach also to the N = 126
r-process waiting points.

\begin{figure}
\begin{center}
\includegraphics[width=0.95\linewidth]{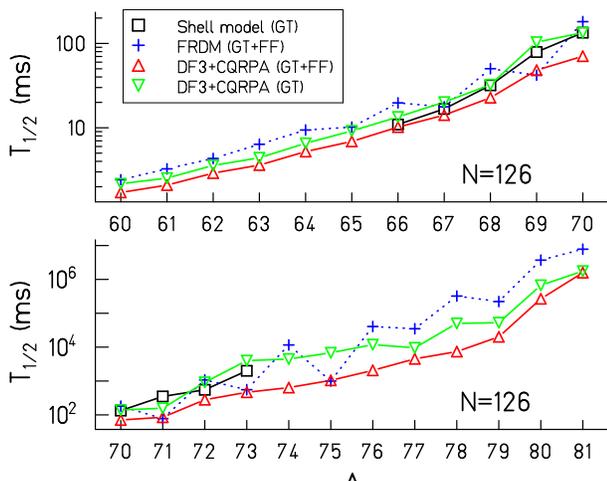}
\vspace{-8mm} \caption{(Color online). Calculated total
$\beta$-decay half-lives for $N = 126$
  isotones: triangles correspond to DF3+CQRPA calculations including only
  Gamow-Teller transitions (triangles-down) and Gamow-Teller plus
  first-forbidden transitions (triangles-up); Squares correspond to Shell-Model
  (GT) \cite{rmp} calculations and crosses correspond to FRDM(GT+FF) half-life
  \cite{MPK02} calculations.
\label{T_theor2}}
\end{center}
\end{figure}

The calculated half-lives for the $N=126$ isotones are shown in
Fig. 4.  Compared to the FRDM \cite{MPK02}, which is often used in
r-process simulations, the DF3+QRPA gives substantially shorter
half-lives and does not show a pronounced odd-even staggering.
Thus the break through the $N=126$ waiting points, and hence the
matter flow to the heavier fissioning nuclei \cite{gmp07}, is
faster. The GT contributions to the DF3+QRPA half-lives are very
similar to those calculated with the shell model \cite{rmp} and
somewhat faster than the HFB results \cite{Engel99}. We note,
however, that FF transitions, which speed up the half-lives (Fig.
4), are missed in the calculations of \cite{rmp,Engel99}.

In summary, an important step has been achieved in the study of new neutron-rich nuclei approaching
the neutron closed shell at N=126.  It has been shown that the N=126 region far below the doubly
magic $^{208}$Pb has become accessible experimentally, and half-lives for nuclei reasonably close
to the $N=126$ r-process waiting points have been determined. Thus, for the first time data exist
which put stringent constraints on the theoretical models required to calculate the half-lives of
these r-process nuclei. We find that a comparison of the measured half-lives with those of a
DF+QRPA calculation presented here is quite satisfactory, placing some confidence into the prediction
of this model for the half-lives of the $N=126$ r-process waiting-point nuclei.  The calculations
indicate significant contributions of first-forbidden transitions to the half-lives which have been
either ignored or insufficiently accounted for in previous studies.

\begin{acknowledgments}
This work was supported by the Spanish MEC (FPA2005-00732), Xunta
de Galicia (Conseller\'{\i}a de Educaci\'on) and by the European
Commission (EURONS, contract no. 506065). I.N.B acknowledges the
support by German DFG under the contract 436 RUS 113 907/0-1.
T.K.-N. acknowledges the support from Xunta de Galicia,
Direcci\'on Xeral de Investigaci\'on, Desenvolvemento e
Innovaci\'on (Bolsa Predoutoral, Conseller\'{\i}a de Innovaci\'on
e Industria)  and Universidade de Santiago de Compostela, through
the PhD studies.
\end{acknowledgments}


\end{document}